\documentclass[12pt]{article}

\usepackage{amsmath}
\usepackage[usenames]{color}
\usepackage{graphics}
\usepackage{hangcaption}

\allowdisplaybreaks

\begin{document}

\baselineskip=18.6pt plus 0.2pt minus 0.1pt

\makeatletter
\@addtoreset{equation}{section}
\renewcommand{\theequation}{\thesection.\arabic{equation}}


\newcommand{\der}[1]{\partial_{#1}}
\newcommand{\dder}[2]{\frac{\partial{#1}}{\partial{#2}}}
\newcommand{\inv}[1]{\frac{1}{#1}}
\newcommand{\Rint}[1]{\int \hspace{-0.2em} {\rm d}{#1}\;}
\newcommand{\Tr}{{\rm Tr}}
\newcommand{\tr}{{\rm tr}}

\newcommand{\ph}[1]{\phantom{#1}}
\newcommand{\nn}{\nonumber}

\newcommand{\pref}[1]{(\ref{#1})}
\newcommand{\vbar}[1][]{\Bigr|_{#1}}
\newcommand{\tm}{\mbox{$\times$}}
\newcommand{\gf}[1]{\langle \; #1 \; \rangle}
\newcommand{\uslash}{\slash\hspace{-0.55em}u}
\newcommand{\graph}[2]{\scalebox{#1}{\includegraphics{#2}}}
\makeatother

\begin{titlepage}
\title{
\hfill\parbox{4cm}
{\normalsize KUNS-1813\\{\tt hep-th/0211272}}\\
\vspace{1cm}
Improved Perturbation Theory \\[5pt]and\\[5pt]
Four-Dimensional Space-Time in the IIB Matrix Model
}

\author{
{\sc Hikaru}~{\sc Kawai}\thanks{{\tt hkawai@gauge.scphys.kyoto-u.ac.jp}}
,\hspace{5pt}
{\sc Shoichi}~{\sc Kawamoto}\thanks{{\tt kawamoto@gauge.scphys.kyoto-u.ac.jp}}
,\hspace{5pt}
{\sc Tsunehide}~{\sc Kuroki}\thanks{{\tt kuroki@gauge.scphys.kyoto-u.ac.jp}}
\\ {}\hspace*{2pt} and \hspace*{0pt}
{\sc Shun'ichi}~{\sc Shinohara}\thanks{
{\tt shunichi@gauge.scphys.kyoto-u.ac.jp}}
\\[7pt]
{\it Department of Physics, Kyoto University, Kyoto 606-8502, Japan}
}

\date{\normalsize November, 2002}
\maketitle
\thispagestyle{empty}
\begin{abstract}
\normalsize\noindent
We have analyzed the IIB matrix model on the basis 
of the improved mean field approximation (IMFA) and have obtained 
evidence suggesting that the four-dimensional space-time appears 
as its most stable vacuum.
This method is a systematic way to obtain an improved perturbation series and
 was first applied to the IIB matrix model by Nishimura and Sugino.
In a previous paper, we reformulated this method and proposed a criterion
for the convergence of the improved series, that is, the appearance of a
``plateau.''
In this paper, we carry out higher-order calculations, and find that our
improved free energy tends to have a plateau, which shows that IMFA
works well in the IIB matrix model.
\end{abstract}

\end{titlepage}

\section{Introduction and summary}
\label{sec:intro_summry}

\vspace*{0pt}

String theory has been proposed as a unified theory of fundamental
interactions including gravity. If it is the well-defined quantum
gravity theory which our nature adopts, it should be able to predict
several properties of our universe, for instance, the gauge symmetry at low
energy, the particle content and their masses, and the dimensionality
of our space-time.
The last of these is the property we would like to investigate
in this paper.

As is well known, superstring theory has infinitely many vacua with
various dimensionalities that arise perturbatively.
One of the most promising scenarios to construct a realistic
four-dimensional model from superstring theory is a compactification in
which the ten-dimensional space-time consists of a flat
four-dimensional space-time and a six-dimensional compact manifold
that is small enough to be invisible to our experiments.
There are several ways to realize this scenario, for example,
 Calabi-Yau compactification \cite{CY_cpt}, fermionic construction
 \cite{KLT1987}, and so on, but they are all stable and there is no way
 to single out the true vacuum perturbatively.
However, it has been revealed that this problem originates from
perturbative formulations themselves, and to overcome this difficulty,
we need a nonperturbative formulation and analysis of string theory.

In the mid-1990s, some models were presented as a constructive
formulation of M-theory \cite{Matrices} and certain kinds of superstring
theories, such as type IIA matrix string \cite{IIA_matrices}, type I
superstring \cite{typeI_matrices} and heterotic superstring
\cite{heterotic_matrices}. 
Here we would like to analyze the model called the IKKT model
\cite{Ishibashi:1997xs}, which is conjectured to be a nonperturbative
definition of type IIB superstring theory.
It seems to be most promising for nonperturbative analyses of
superstring theory, and some kinds of extensions of
this model have been proposed \cite{new_matrices}.
For a review of the IIB matrix model, see Ref. \cite{Aoki:1999bq}. 
Also, there are some mechanisms proposed for dynamical breakdown of
Lorentz symmetry in this model (see Ref. \cite{Nishimura:2000}).

In order to analyse the IIB matrix model, we use a method that we call
the ``improved mean field approximation'' (IMFA)\cite{KKKMS1}.
This approximation was applied to the large-$N$ reduced
Yang-Mills models by Oda and Sugino \cite{Sugino:2001}. 
Then, Nishimura and Sugino 
applied it to the IIB matrix model in an excellent work
\cite{Nishimura:2001sx}.
They obtained a result that suggests the breakdown of Lorentz
symmetry to the four-dimensional symmetry.
In a previous paper \cite{KKKMS1}, we analyzed how the IMFA scheme works
and discovered a general structure that we call the ``improved Taylor
expansion.'' 
Furthermore, we proposed a principle for choosing the mean
fields, that is, the existence of a ``plateau.'' 
The emergence of a plateau indicates that the approximation scheme
works well and, in fact, its existence can be confirmed 
in some exactly solvable models.
(See Ref. \cite{KKKMS1} for reference. There, many examples are given which
show how good this scheme is.)
Furthermore we developed a computational method using two-particle
irreducible (2PI) graphs, which simplifies the calculation
drastically.
The 2PI free energy has a close relationship with the Schwinger-Dyson
equation as discussed in Ref. \cite{KKKMS1}.

Using this method, we calculated the free energy up to 5th order and
obtained a preliminary result which suggests that the eigenvalue
distribution of the matrices preserves only the four-dimensional
rotational symmetry.

In this paper, we carry out a further calculation up to 7th order. 
We find evidence suggesting the emergence of a plateau, which was not
clear at 5th order.
In fact, we have evaluated the free energy and the extent of the eigenvalue
distribution for various Lorentz symmetries by introducing corresponding
mean fields.
In the case of SO(4) symmetry, the number of extrema of the free
energy increases as we go to higher orders, and it seems that they 
form a plateau.
Furthermore, the eigenvalues are widely
distributed in the four ``non-compact'' directions, 
while they tend to gather in the six compact directions.
One the other hand, if we impose SO(7) symmetry, the number of 
extrema of the free energy does not grow enough to form a plateau, and
the eigenvalues are distributed somewhat isotropically in the ten
directions.
As shown in Ref. \cite{KKKMS1}, the other cases are reduced to the above
two cases, or do not have a plateau at all.

We evaluate the 2738 graphs to obtain the free energy up
to 7th order.
All the calculations, including the generation and
computation of the graphs, are totally automatized now, and we will be able
to go further and find reliable results.

In \S \ref{sec:IMFA}, we provide a short review of the improved
mean field approximation  with a $\phi^4$ matrix model as an example.
We examine a distribution of extrema of the free energy of this
model, which provides a method of the search for a plateau 
in the IIB matrix model.
In \S \ref{sec:FofIIBMM}, we apply IMFA to the IIB matrix model.
We obtain the free energy, investigate the distribution of its
extrema to search for a plateau and examine the eigenvalue distributions.
Section \ref{sec:conlusion} is devoted to a conclusion and discussion.
In Appendix \ref{sec:matrix_QED}, we make use of a $\phi^4$-QED type
matrix model to confirm our counting of the graphs.

\section{Improved Mean Field Approximation for the Free Energy}
\label{sec:IMFA}

In this section, we review the improved mean field approximation that
was developed in our previous paper \cite{KKKMS1}.
Although all the techniques we introduce here are the same as
those used in Ref. \cite{KKKMS1}, we explain them to make this paper
self-contained and for reader's convenience.

\subsection{Ordinary vs. Improved perturbation theory}

Suppose we have some action function $S(x) = S_0(x) + S_1(x)$ and its
free energy
\begin{equation}
  F = - \ln Z = - \ln \left( \Rint{x} e^{-S(x)} \right) ,
\end{equation}
where $S_0(x)$ is the unperturbed part, which can be integrated analytically, 
and $S_1(x)$ is difficult or impossible to integrate and is treated as 
a perturbation of $S_0$.

In ordinary perturbation theory, we keep the unperturbed part in
the exponential and expand the perturbation into a series.
For convenience, we introduce a formal coupling constant $g$ and
rewrite the action as $S = S_0 + g S_1$. Then we obtain the ordinary
perturbation series for the free energy with respect to $g$ as follows:
\begin{equation}
  \label{eq:ord_pert}
  F = - \ln \left( \Rint{x} \sum^{\infty}_{k=0} \frac{(-g)^k}{k!} S_1^k \
  e^{-S_0(x)} \right).
\end{equation}
We can obtain the free energy up to order $n$ in the
perturbation theory  by truncating the infinite series up to order
$n$ and setting $g=1$.

In general, the perturbation series has a finite (or maybe even zero) 
convergence radius with respect to the parameters appearing in the
action, for example, the inverse mass squared $1/m_0^2$ in $S_0$.
To see what happens and when the perturbation theory fails, let us
consider a zero-dimensional $\phi^4$ one-matrix model.

\subsection{The $\phi^4$ matrix model}

The action of the zero dimensional $\phi^4$ matrix model is
\begin{equation}
  \label{eq:ac_phi4}
  S = \frac{m_0^2}{2} \Tr \phi^2 + \frac{1}{4} \Tr \phi^4,
\end{equation}
where $\phi$ is an $N \times N$ hermitian matrix.
In the ordinary perturbation theory,
 we treat the quadratic part, $m_0^2 \phi^2 /2$, as the
unperturbed action $S_0$ and the $\phi^4 /4$ part as the perturbation
$S_1$.
Here, as an example, we consider the free energy. 
After introducing a formal coupling constant $g$  and expanding the
 exponential part with respect to it, we obtain
\begin{equation}
  \label{eq:phi4_pert}
  F= - \ln \left[ \sum^{\infty}_{k=0} \Rint{\phi} \left( \frac{-g}{4}
  \right)^k \frac{(\Tr \phi^4)^k}{k!} \exp \left(- \frac{m_0^2}{2} \Tr \phi^2
  \right) \right], 
\end{equation}
and we can estimate this series using the ordinary Feynman diagram
technique.
Here we consider the large-$N$ limit, where only
planar graphs contribute to the free energy. 
Thus, in the ordinary perturbation theory,
the free energy is given by a series with respect to $g/m_0^4$ as
follows:
\begin{align}
  \label{F_phi4}
F = & \, \, \raisebox{- 6pt}{\scalebox{0.5}{\includegraphics{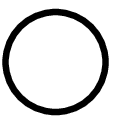}}} + 
       \raisebox{-13pt}{\scalebox{0.5}{\includegraphics{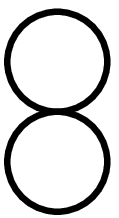}}} + 
       \raisebox{-17pt}{\scalebox{0.5}{\includegraphics{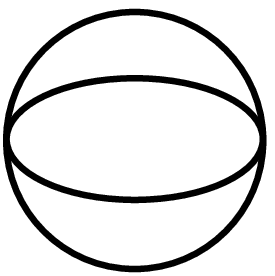}}} +
       \raisebox{-22pt}{\scalebox{0.5}{\includegraphics{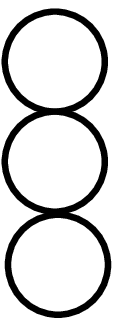}}} +
       \raisebox{-24pt}{\scalebox{0.5}{\includegraphics{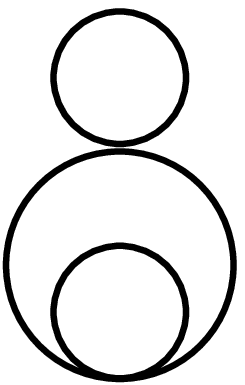}}} +
       \mathcal{O}(g^3) \nn\\
 = & -\frac{1}{2} \ln \left( \frac{1}{m_0^2} \right) + \frac{1}{2}
       \frac{g}{m_0^4} - \frac{9}{8}\frac{g^2}{m_0^8} +
       \mathcal{O}(g^3) .
\end{align}
On the other hand, one can evaluate this free energy exactly by
analyzing the eigenvalue distribution \cite{Brezin:1978sv}, 
and it is known that the convergence radius of this series is
$1/12$. Thus, after we set the formal coupling  $g$ to 1, this series
converges when $m_0^4 > 12$.
This means that we cannot calculate the free energy for the massless case,
i.e. $m_0^2=0$, as the limit of this perturbation series.
Because the IIB matrix model of interest does not have a quadratic
term, it corresponds to the massless case, in which the ordinary
perturbation theory does not work.

To overcome this difficulty we introduce a new method, the
improved mean field approximation, and obtain an improved perturbation
series.

\subsection{Improved mean field approximation}

When we apply the improved mean field approximation scheme, we first
introduce a ``mean field'' $S_{\rm m}(x,a)$ which can be easily
integrated, for instance, a quadratic term.
Here, $a$ is a set of parameters in the mean field, and we will 
tune it later to make the approximation better.
Then, we rewrite the original action as 
\begin{equation}
  S = S_0 + S_1 \ \Rightarrow \ S = S_{\rm m} + (S_0 + S_1
  - S_{\rm m} ),
\end{equation}
and we take $S_{\rm m}$ as an unperturbed action and the part within
parenthesis as the perturbation. 
We introduce a formal coupling constant $g$ as before. 
Thus the action becomes $S=S_{\rm m} + g(S_0 + S_1  - S_{\rm m} )$, and
the expansion of the exponential with respect to $g$ yields another
perturbation series, which we call the improved perturbation series.
Finally, we tune the set of parameters $a$ to make the improved series
converge, as we explain below.
We call this procedure the ``improved mean field approximation.''
In particular, if we consider the first order of this approximation
for the free energy defined in (\ref{F_phi4}), and tune the parameter
$a$ using the condition $dF /d a =0$,  it is simply the ordinary
mean field approximation.

As an example, we consider again the zero-dimensional $\phi^4$
matrix model.

\subsection{IMFA for the $\phi^4$ matrix model}
\label{sec:IMFA_phi4}

In this case, we consider a quadratic term as the mean field:
\begin{equation}
  S_{\rm m} = \frac{m^2}{2} \Tr \phi^2 .
\end{equation}
Then, we construct a modified action, including a formal coupling $g$, 
as 
\begin{align}
\label{eq:mod_phi4}
  S =& \frac{m^2}{2} \Tr \phi^2 + g \left( \frac{m_0^2}{2} \Tr \phi^2 +
  \frac{1}{4} \Tr \phi^4 - \frac{m^2}{2} \Tr \phi^2 \right)  \nn\\
=& \frac{g}{4} \Tr \phi^4 + \frac{m^2 + g(m_0^2 - m^2)}{2} \Tr \phi^2 .
\end{align}
We emphasize that we do not need to calculate graphs again.
By comparing (\ref{eq:mod_phi4}) with the original action
(\ref{eq:ac_phi4}), we have only to substitute $m^2 +g(m_0^2 - m^2)$
for the mass squared $m_0^2$ appearing in the ordinary perturbation
theory.

Therefore, by substituting  $m^2 + g(m_0^2 - m^2)$ for $m_0^2$, and
re-expanding with respect to $g$, (\ref{F_phi4}) becomes 
\begin{align}
  \label{eq:F_imp_phi4}
  F_{\rm improved} =& -\frac{1}{2} \ln \left( \frac{1}{m^2 + g(m_0^2 -
  m^2)} \right) + \frac{g}{2} \frac{1}{(m^2 + g(m_0^2 - m^2))^2} \nn\\
  & \phantom{aaaa} - \frac{9}{8} g^2 \frac{1}{(m^2 + g(m_0^2 - m^2))^4} +
  {\cal O}(g^3) \nn\\
=& -\frac{1}{2} \ln \left( \frac{1}{m^2} \right) 
+ \frac{g}{2} \left( \frac{1}{m^4} + \frac{m_0^2 - m^2}{m^2}\right)
  \nn\\
& \phantom{aaaa} + g^2 \left( -\frac{9}{8}\frac{1}{m^8} -
  \frac{m_0^2-m^2}{m^6}  -\frac{1}{4}\frac{(m_0^2 - m^2)^2}{m^4}
  \right) + {\cal O}(g^3).
\end{align}

In this form, the massless limit is no longer singular and can be
taken simply by setting $m_0^2=0$.
Furthermore, from this form, we know how to calculate 
the improved perturbation series even for a massless theory, 
to which the ordinary perturbation theory cannot be applied.
First, we add a mass term $m^2 \phi^2 /2$ to the original action by
hand and calculate the free energy perturbatively.
Then we substitute $m^2 - g m^2$ for  $m^2$ and re-expand it with
respect to $g$.
Finally, by setting $g=1$, we obtain the improved perturbation series
for the massless theory.
As we see, this prescription is useful for calculating the IIB matrix
model, which also has no quadratic term.

\begin{figure}[htbp]
  \begin{center}
    \leavevmode
    \begin{tabular}{cc}
    \scalebox{0.35}{\rotatebox{-90}{{\includegraphics{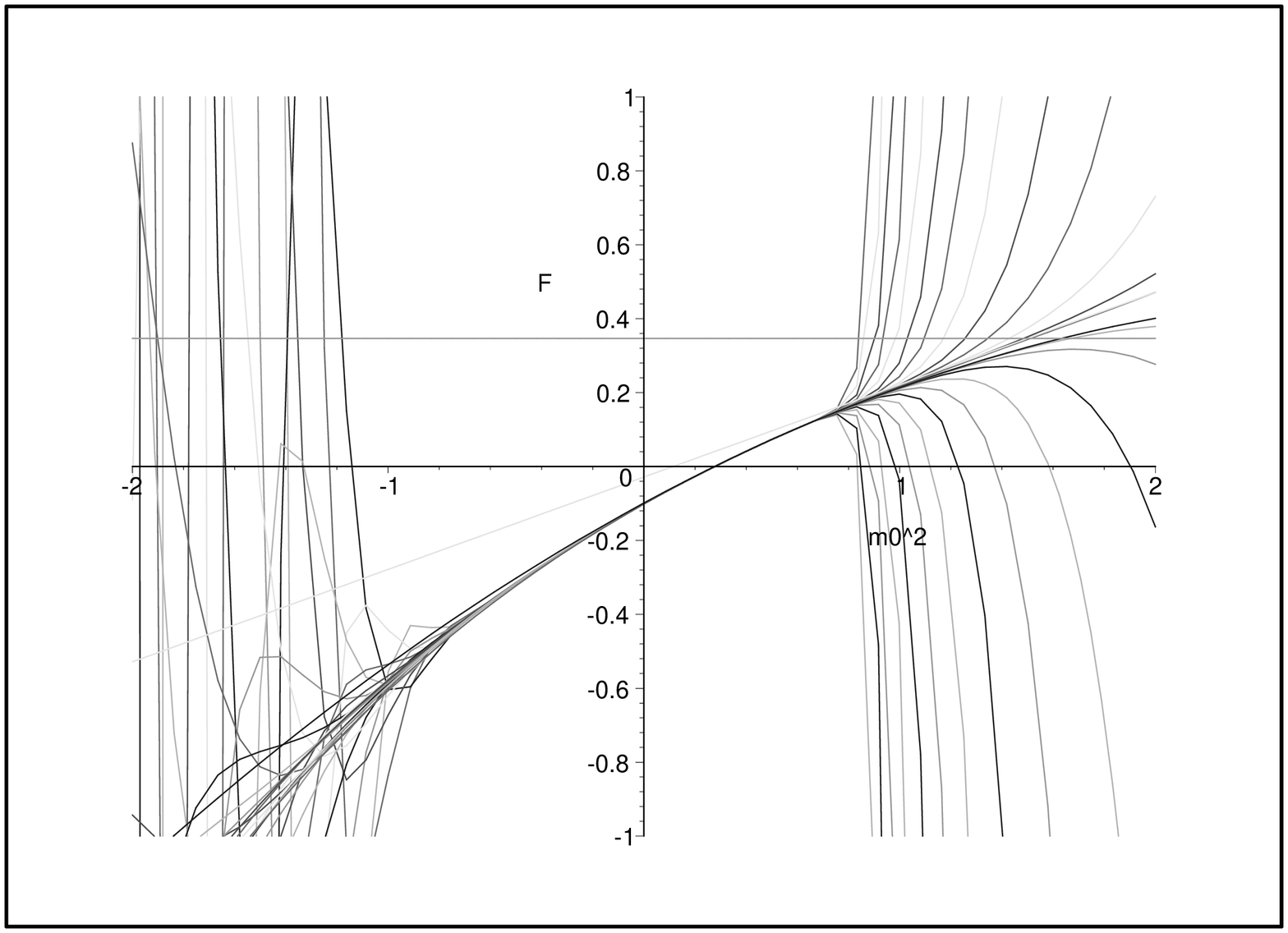}}}} &
    \scalebox{0.35}{\rotatebox{-90}{{\includegraphics{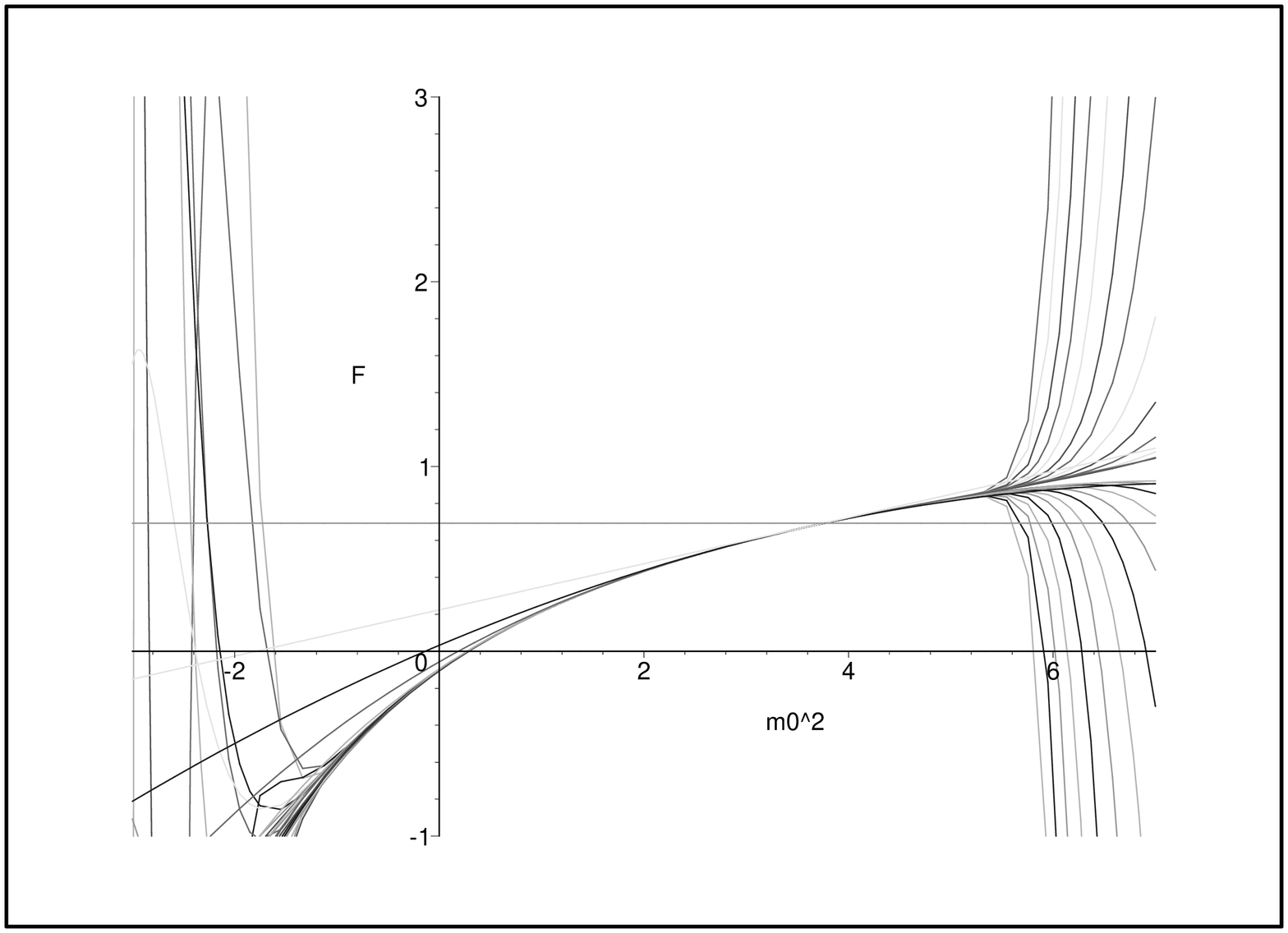}}}} \\
    $m^2=2$ & $m^2=4$ \\
    \end{tabular}
    \captionwidth = 30em
    \hangcaption{Free energies with $m^2=2$ (left) and
    $m^2=4$ (right).
       The horizontal axis denotes $m_0^2$.}
    \label{fig:phi4_m=2and4}
  \end{center}
\end{figure}

Let us return to the case of a general value of $m_0^2$.
The final step of this prescription is to tune the parameter $m^2$ so
that the improved series converges. 
To elucidate this situation, in Fig. \ref{fig:phi4_m=2and4}, we plot
the improved free energies with respect to $m_0^2$ at various orders for
$m^2=2$ and $4$.
It is clear that the improved series converges quite well in some domain. 
The position of the domain of the convergence
depends on the choice of the parameter $m^2$, for example, around
$m_0^2=0$ for $m^2=2$ and around $m_0^2=4$ for $m^2=4$.
Here,
we have first plotted the improved series with respect to $m_0^2$ for various
values of $m^2$, and
then determined the value of $m^2$ to make the series converge 
around the value of $m_0^2$ in question.

\subsection{The plateau and how we identify it}

\begin{figure}[htbp]
  \begin{center}
    \leavevmode
  \begin{tabular}{cc}
  \scalebox{0.3}{\rotatebox{-90}{{\includegraphics{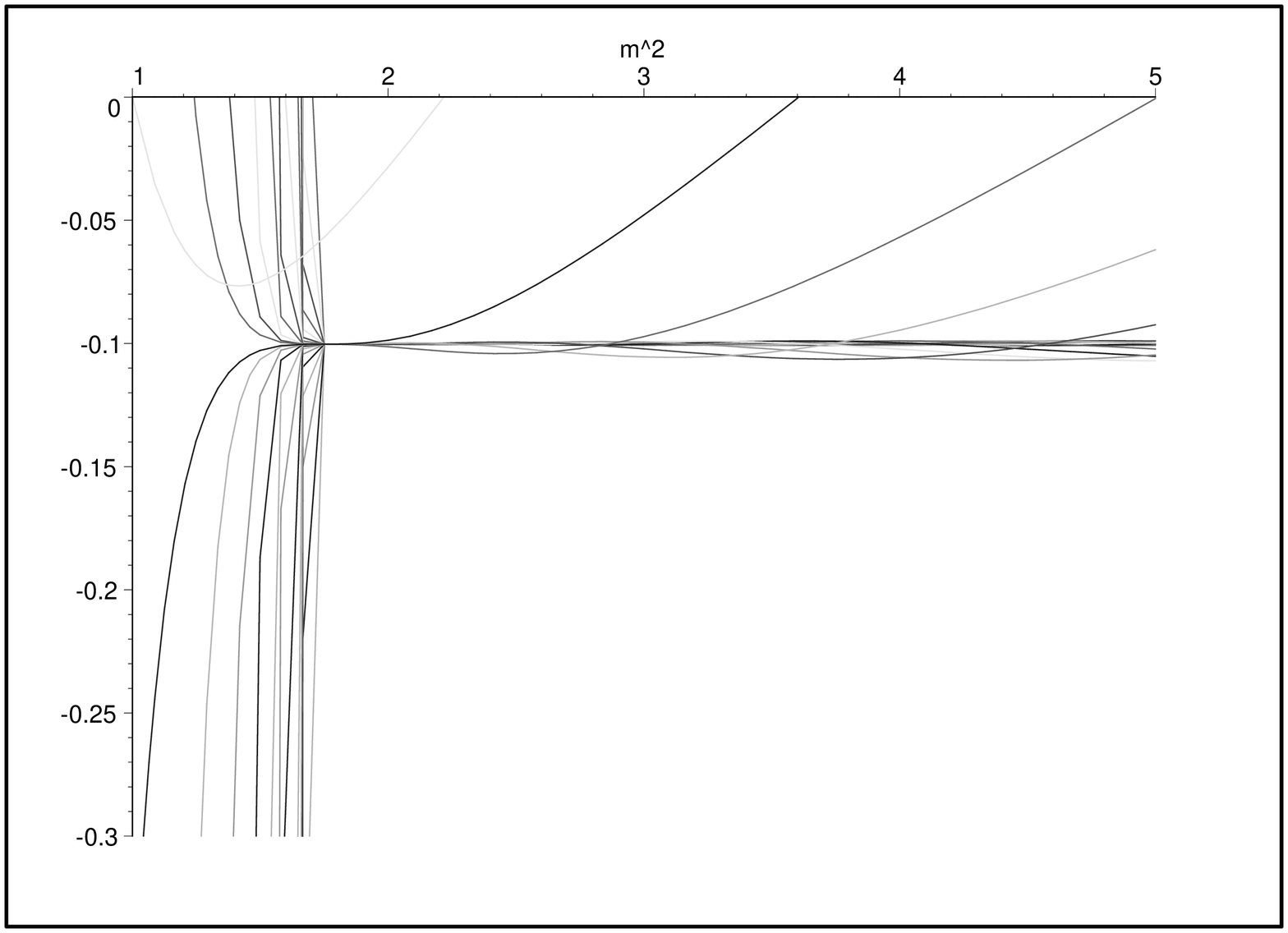}}}} &
  \scalebox{0.3}{\rotatebox{-90}{{\includegraphics{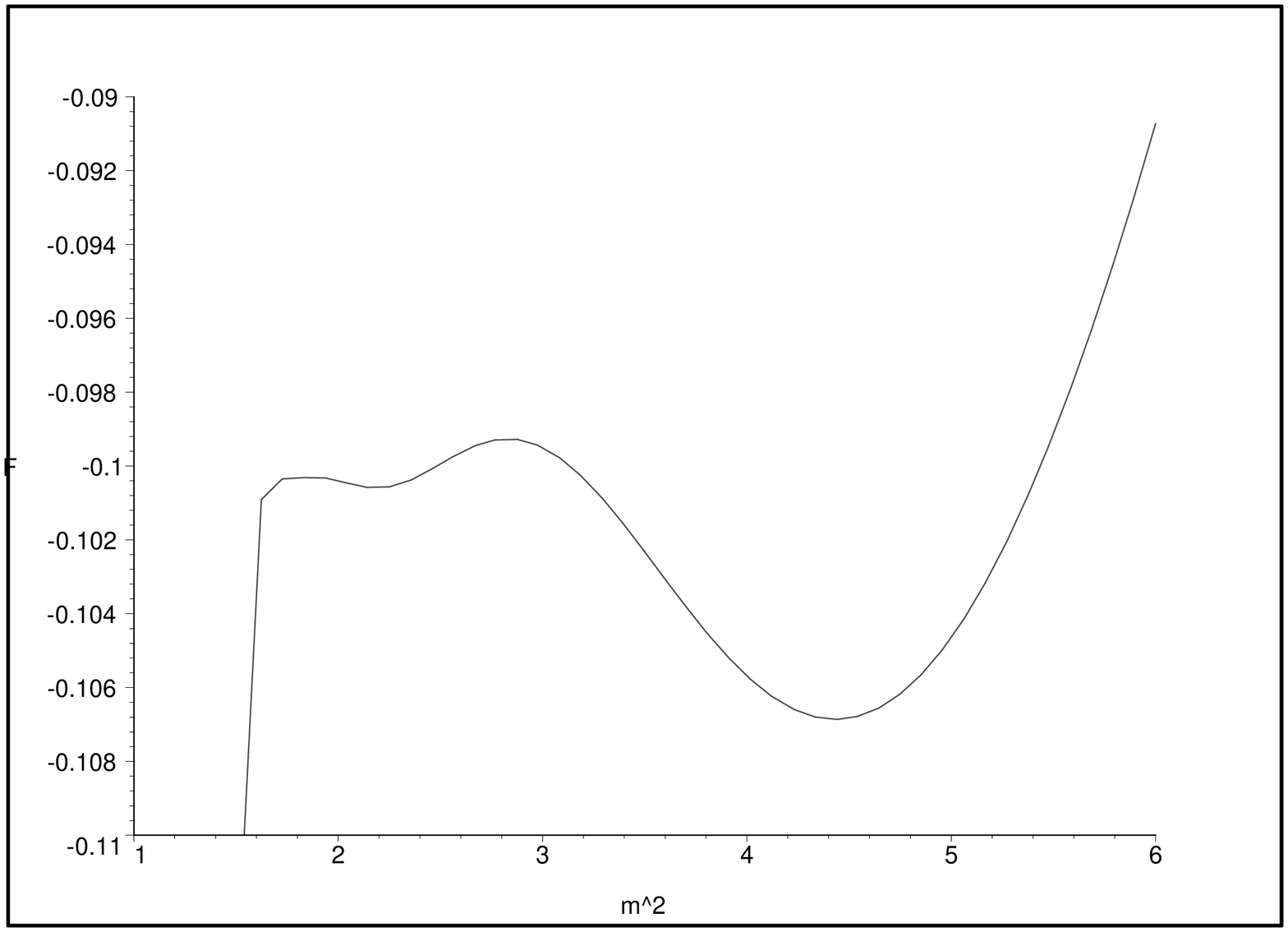}}}} \\
   From 0th to 29th order & 7th order \\
  \scalebox{0.3}{\rotatebox{-90}{{\includegraphics{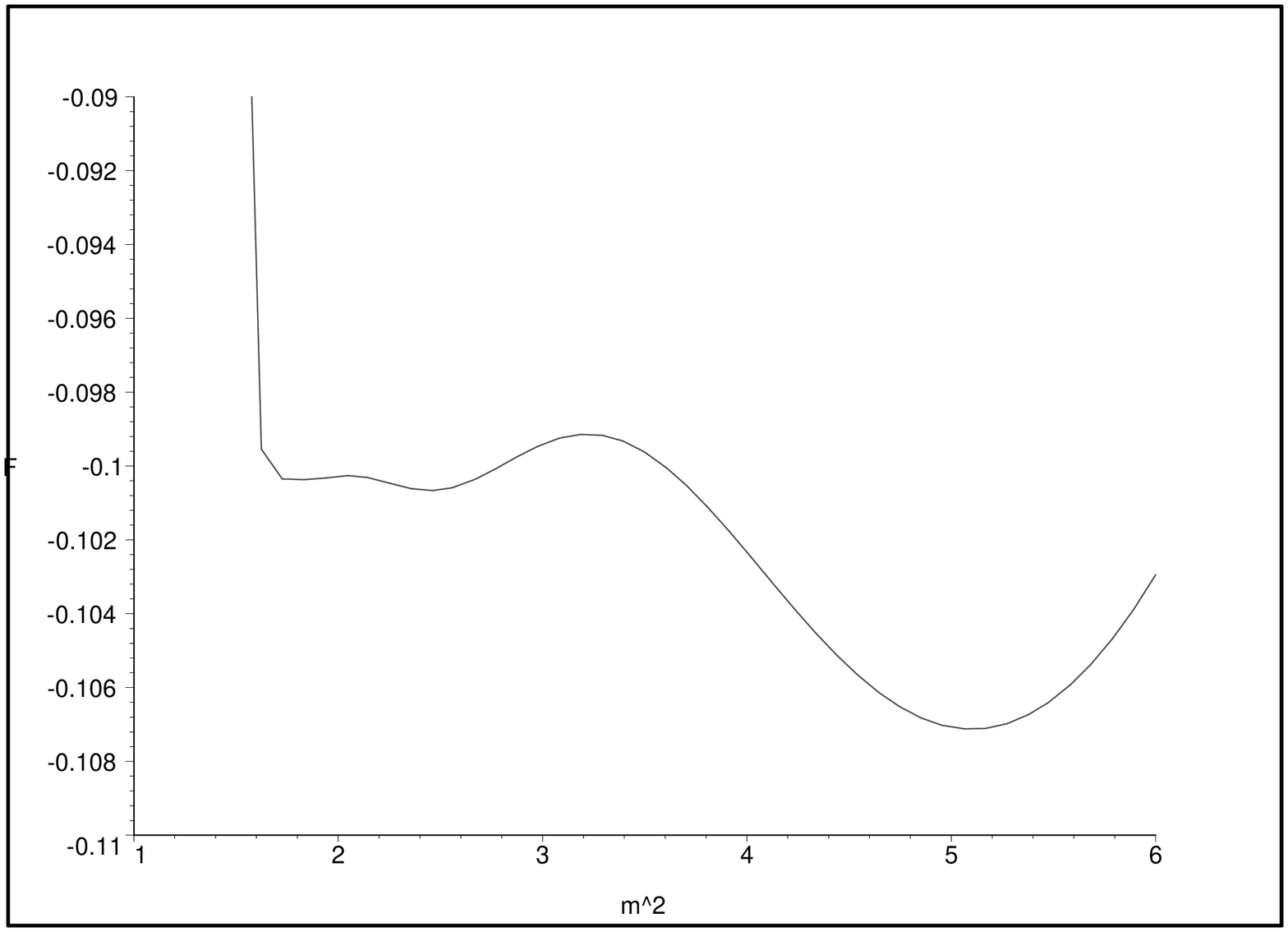}}}} &
  \scalebox{0.3}{\rotatebox{-90}{{\includegraphics{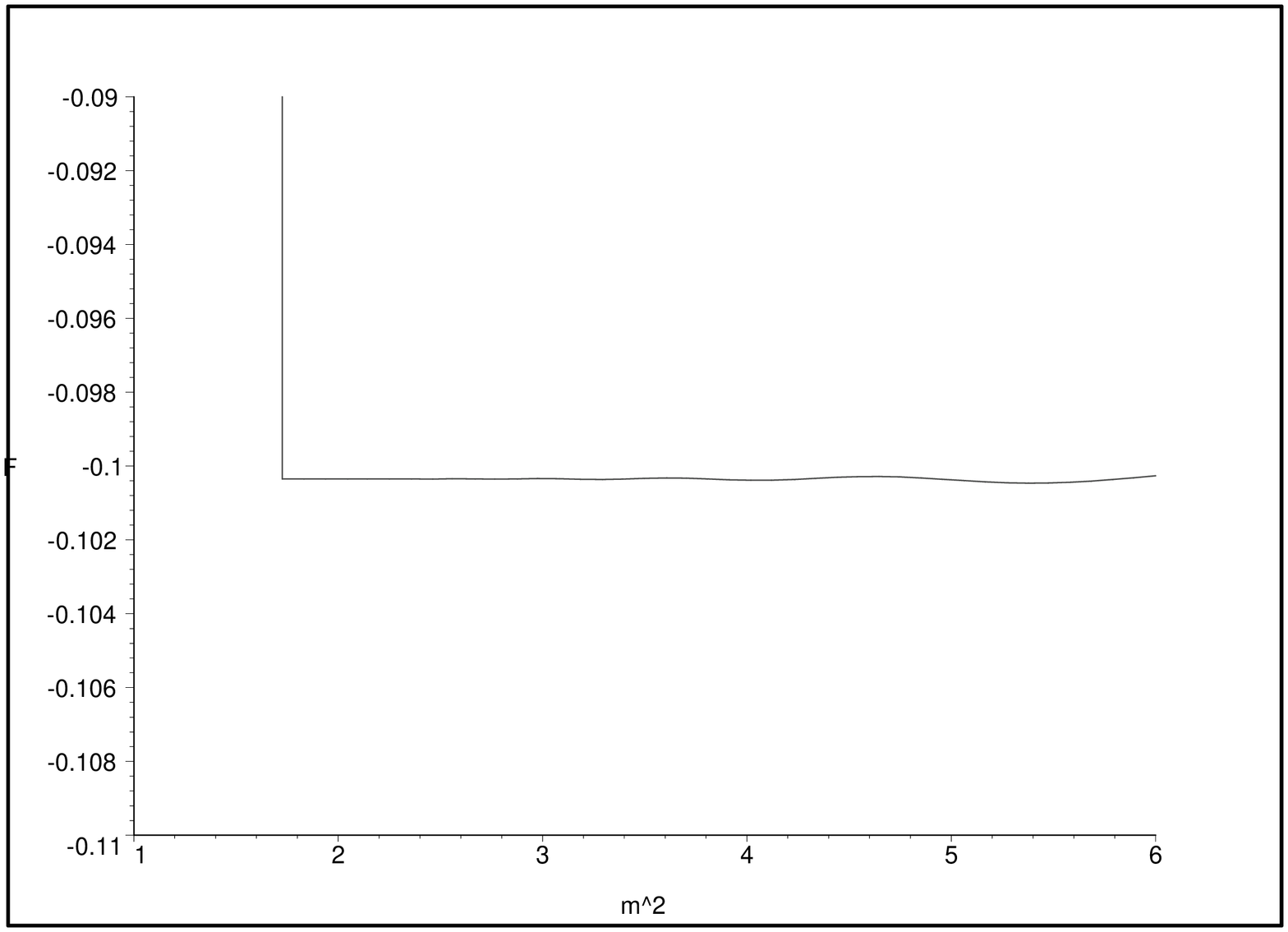}}}} \\
   8th order & 27th order \\
  \end{tabular}
  \captionwidth = 30em
    \hangcaption{Free energies with $m_0^2=0$ at 7th, 8th, 27th order
    and from 0th to 29th order.
       The horizontal axis denotes $m^2$.}
    \label{fig:phi4_plateau}
  \end{center}
\end{figure}

There is an easier way to find an appropriate value of the
parameter $m^2$ than that considered above for a fixed value 
of the bare parameter $m_0^2$, for
example, the massless case $m_0^2=0$.
Fig. \ref{fig:phi4_plateau} displays the improved free energies as
functions of $m^2$ with $m_0^2=0$. 
We see that a plateau develops at higher orders.
Intuitively, this fact is very natural, because $m^2$ is an artificially
introduced parameter, and the true value of the free energy must be
independent of its choice.
Some years ago, Dhar and Stevenson advocated a ``principle of
minimal sensitivity'' \cite{Stevenson:1981vj}.
They stated that in the improved perturbation theory one should choose
the set of parameters such that the improved quantity is stationary
with respect to it, i.e. $\partial F_{\rm improved} / \partial m^2 =0$
in our case.
However, we claim that the criterion for good approximation should be
the existence of a plateau; that is, one should choose the set of
parameters to be on the plateau.
This claim was first made in a previous paper \cite{KKKMS1},
 and some people have been studying the properties of the plateau and seeking
 a better definition, especially in the case that there are many parameters
 in the mean fields.
In such cases, it becomes difficult to identify a plateau, because we
cannot visualize it easily.\footnote{%
Nishimura, Okubo and Sugino proposed  ``the histogram prescription'' to
identify a plateau for the many parameter case
\cite{Nishimura-Okubo-Sugino}.
}

\begin{figure}[htbp]
  \begin{center}
    \leavevmode
  \scalebox{1.1}{\includegraphics{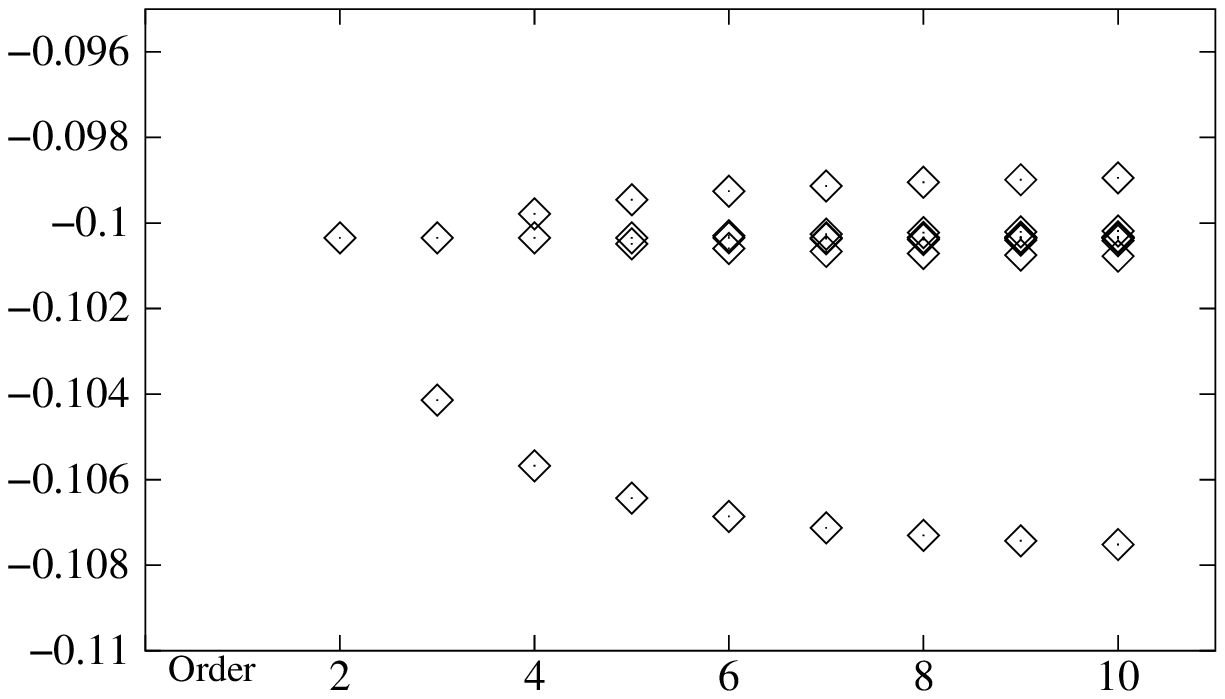}}
  \captionwidth = 30em
    \hangcaption{Extrema of the improved free energy of the massless $\phi^4$
  matrix model from 1st to 10th order.}
    \label{fig:phi4_extrema}
  \end{center}
\end{figure}

To obtain a concrete criterion for the plateau, we plot the extrema
of the improved free energy of the massless $\phi^4$ matrix model with
respect to $m^2$ in Fig. \ref{fig:phi4_extrema}.
It can be seen from this figure that there are
two characteristic extrema up to 10th order, except at the two
lowest orders.
At each order, one of the two extrema gives the highest free energy, 
which is quite stable, and the other gives the lowest free energy, 
which decreases as we go to higher orders.
Hereafter we call the highest and the lowest extremum the
``overshooting'' and ``undershooting'' extremum, respectively.
The other extrema tend to accumulate between these two extrema, 
and they are expected to form a plateau.
In fact, if we take a closer look at a plateau, we find that it
consists of many extrema.
In other words, the accumulation of extrema leads to a plateau.
In general, at much higher orders, a finite number of extrema yield 
higher or lower free energies.
However, the situation remains unchanged in which many of the other extrema
accumulate between these higher and lower extrema and contribute to
the formation of the plateau.


This is the key observation for finding a plateau presented in this paper.
In the next section, we explore the extrema of the free energy of the IIB
matrix model and see whether they accumulate to form a plateau or not.

\section{Free Energies of IIB Matrix Model}
\label{sec:FofIIBMM}

In this section, we would like to analyze the IIB matrix model
\cite{Ishibashi:1997xs}. The action is given by
\begin{equation}
  \label{eq:ac_IIB}
   S_{\rm IIB} = - \Tr \Bigl(\; \frac{g_0^2}{4} [A_\mu,A_\nu]^2
      -\frac{g_0}{2}\, \bar{\psi} \Gamma^\mu [A_\mu, \psi] \;\Bigr) ,
\end{equation}
where $A_\mu~(\mu =1,\cdots,10)$ and $\psi^\alpha~(\alpha=1,\cdots,16)$
are all $N \times N$ hermitian matrices transforming as the vector and
left-handed spinor representations under SO(10).
$g_0$ is the only parameter in this model, which is dimensional, and 
 we take the large $N$ limit while fixing the 't Hooft coupling
$g_0^2N$ to $1$.
This is defined as a zero-dimensional reduced model of a ten-dimensional
super Yang-Mills theory.
Because it has no quadratic term, we cannot apply the ordinary
perturbation theory.

We use the improved mean field approximation to evaluate the free 
energy.
Following the method explained in the last section, we add and
subtract a quadratic term for bosonic and fermionic matrices as mean
fields.
Then our action becomes
\begin{align}
\label{IIB_action}
  S =&  - N \  \Tr \Bigl(\; \frac{g}{4} [A_\mu,A_\nu]^2
      -\frac{\sqrt{g}}{2}\, \bar{\psi} \Gamma^\mu [A_\mu, \psi]
      \;\Bigr)  + S_{\rm m} -g S_{\rm m}, \\
 S_{\rm m} =& \frac{1}{2} \, (C^{-1})^{(\mu\nu)} N\, \Tr
  (A_{\mu} A_{\nu} ) -\frac{i}{2} \left( {\cal C} \uslash ^{-1}
  \right)_{[\alpha \beta]} N\, \Tr \left( \psi^{\alpha} \psi^{\beta}
  \right),
\end{align}
where $S_{\rm m}$ is the quadratic term introduced 
as the mean fields\footnote{
${\cal C}$ is the charge conjugation matrix defined as
$ {}^t{\cal C}=-{\cal C}, \hspace{1em}
  {\cal C}\Gamma^{\mu} = - {}^t \Gamma^{\mu} {\cal C}
    \hspace{0.5em} ( \mu =1,\cdots ,10 )$.}.
Here we introduce the formal coupling constant $g$ as $g_0^2$ in
(\ref{eq:ac_IIB}).
The coefficients $C_{\mu\nu}$ and 
$\uslash = u_{\mu\nu\rho} \Gamma^{\mu\nu\rho} /3!$ 
are the propagators for bosonic and fermionic matrices in the
perturbation theory, respectively.
$C_{(\mu\nu)}$ is a second rank symmetric tensor, and
$u_{[\mu\nu\rho]}$ is a third rank antisymmetric tensor.

Here we should comment on how these mean fields are constructed.
The symmetries of the original IIB matrix model are the matrix rotation
U($N$), ten-dimensional Lorentz symmetry SO(10), translational
symmetry, which changes $A_{\mu}$ to $A_{\mu} + {\rm const.} \times
\textbf{1}_{N}$, and the type IIB supersymmetry.
Our mean field preserves U($N$), while it breaks SO(10) and supersymmetry.
One might worry that this leads to an inconsistency because the existence
of the type IIB supersymmetry plays an important role in the IIB matrix model.
However, if the supersymmetry is restored in the true ground state of our 
model, the parameters will go to $C=\infty$ and $u=\infty$ and our 
mean field should vanish.
This is the standard story of models in which dynamical
symmetry breakdown occurs, as in the Nambu--Jona-Lasinio model.
In this case, even if we introduce a mass term that breaks the chiral
symmetry, it vanishes in the phase where the symmetry is restored.
If $C$ and $u$ are still finite after the large
$N$ limit is taken, both the Lorentz symmetry and supersymmetry are
spontaneously broken. 
This is the very scenario we expect. 

Here we summarize the procedure we carry out below.
Following the prescription for the massless case described in \S 
\ref{sec:IMFA_phi4}, we first calculate the free energy perturbatively
for the action
\begin{align}
\label{IIB_action_mod}
  S' =&  - N \,  \Tr \Bigl(\; \frac{g}{4} [A_\mu,A_\nu]^2
      -\frac{\sqrt{g}}{2}\, \bar{\psi} \Gamma^\mu [A_\mu, \psi]
      \;\Bigr)  + S_{\rm m} . 
\end{align}
Then we replace $C^{-1}$ and $u^{-1}$ with $(1-g) C^{-1}$ and
$(1-g) u^{-1}$, respectively, in order to recover the contributions 
from the $-g S_{\rm m}$ term.
Finally, we obtain the improved free energy by expanding the result with
respect to $g$.

\subsection{The 2PI free energy and ansatz}

To obtain the free energy, we only have to calculate 
planar connected vacuum graphs. 
However, in such calculations, we have to treat too many graphs at
higher orders.
In order to avoid this problem, we introduce a two-particle irreducible
(2PI) free energy.
The 2PI free energy is considered in Ref. \cite{KKKMS1}
 and it is deeply related to the Schwinger-Dyson equations, as discussed
 in Ref. \cite{KKKMS1}.
In this paper, however, we use the 2PI free energy only as a tool 
to obtain the ordinary free energy easily. {}From this viewpoint, 
the definition and properties of the 2PI free energy are summarized 
as follows:
\begin{itemize} 
\item The 2PI graph is the graph that contains no self-energy graphs as
  its subgraphs; that is, it is two-particle irreducible.
  This means that propagators in a 2PI graph can be regarded as the
  exact propagators.
\item Suppose $G$ is a sum of planar, vacuum, connected 2PI graphs in
  some theory. 
  Then the ordinary free energy in the planar limit is given by
  the Legendre transformation of $G$ with respect to the exact
  propagators.
\end{itemize}
In the IIB matrix model, we also force $G$ to contain no tadpole
graphs, because all the one-point functions satisfy $\gf{A_{\mu}}=0$, 
due to Lorentz invariance.

Once we obtain a 2PI free energy $G$ of (\ref{IIB_action_mod}) up to some
order, the next task is to perform a Legendre transformation.
At this stage, however, we face the new problem that there are too
many parameters to carry out the transformation.
Therefore we need some ansatz that reduces the number of parameters to a
tractable one. 
We use the same ansatz as in Ref. 
\cite{KKKMS1}; that is, we assume several unbroken Lorentz
symmetries that restrict the mean field parameters $C_{\mu\nu}$ and
$u_{\mu\nu\rho}$.
Two peculiar examples, which we consider in this paper, are
the SO(7) $\times$ SO(3) case and the SO(4) $\times$ SO(3) $\times$ SO(3)
$\times$ $Z_2$ case.
As we comment below, these two ansatz are of particular importance
among other possibilities.
When the unbroken symmetry is SO(7) $\times$ SO(3), which we call the SO(7)
ansatz, the parameters associated with the bosonic fields $A_{\mu}$ are
limited as $C_{\mu\nu}={\rm
  diag}(V_1,V_1,V_1,V_1,V_1,V_1,V_1,V_2,V_2,V_2)$, and those with the 
fermionic fields $\psi_{\alpha}$ are restricted according to $u_{8,9,10} =
-u_{9,8,10} = ({\rm cyclic}) = u$, while the others are zero.
The other one is called the SO(4) ansatz, which preserves SO(4) $\times$ SO(3)
$\times$ SO(3) $\times$ $Z_2$ symmetry.
The SO(4) symmetry acts on the indices $\mu=1, \cdots , 4$, and the two SO(3) 
symmetries act on $\mu=5,6,7$ and $\mu=8,9,10$, respectively. 
The $Z_2$ symmetry exchanges $\mu=5,6,7$ and $\mu=8,9,10$
directions, that reverses the 1st direction so that it is an element 
of the SO(10).

Then, in the SO(4) ansatz, the parameters are limited as \\
$C_{\mu\nu}={\rm diag}(V_1,V_1,V_1,V_1,V_2,V_2,V_2,V_2,V_2,V_2)$,
$u_{5,6,7} = u_{8,9,10} = u/ \sqrt{2}$, up to the cyclic permutation of
indices with signature, while the other components are zero.

Here we should comment on the other ansatz.
In Ref. \cite{KKKMS1}, we consider, besides the ansatz mentioned
above, SO(1), SO(2), SO(3), SO(5) and SO(6) ansatz.
For each ansatz, SO(n) represents the residual symmetry for the expanded
direction, which at the end should be understood as the space-time
dimension.
According to the analysis given in Ref. \cite{KKKMS1}, the SO(2) 
and SO(3) ansatz
restore SO(4) symmetry, the SO(5) and SO(6) ansatz restore SO(7)
symmetry, and the SO(1) ansatz has no extrema and, of course, no plateau. 
(See Ref. \cite{KKKMS1} for details.) 
After all, we concentrate on the SO(4)
and SO(7) cases, which behave quite differently, as we now see.

\subsection{Calculation of the free energy and the extent of space-time}
\label{sec:calc}

Now we have all the tools needed to compute the improved free energy.
Our first task is to calculate the 2PI free energy $G$ for the action
(\ref{IIB_action_mod}).
It can be computed as 
\begin{align}
\hspace{-2cm}
\frac{G}{N^2} =& \, \, 
      \raisebox{-13pt}{\scalebox{0.4}{\includegraphics{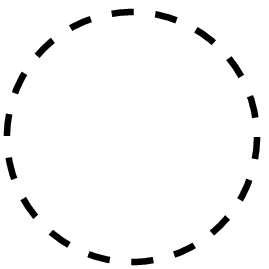}}}
      + \raisebox{-13pt}{\scalebox{0.4}{\includegraphics{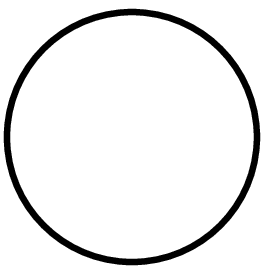}}}
      +g\ \raisebox{-13pt}{\scalebox{0.4}{\includegraphics{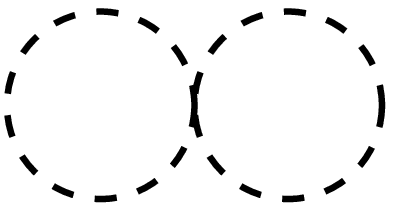}}}
      +g\ \raisebox{-13pt}{\scalebox{0.4}{\includegraphics{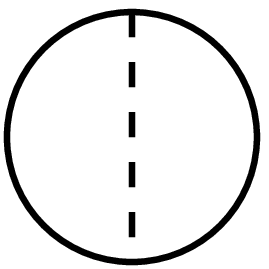}}}
      +g^2\ \raisebox{-13pt}{\scalebox{0.4}{\includegraphics{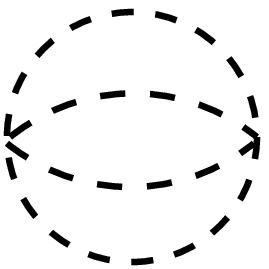}}}
      +g^2\ \raisebox{-13pt}{\scalebox{0.4}{\includegraphics{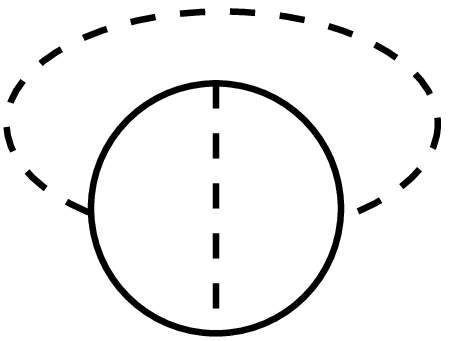}}}
      + {\cal O}(g^3) \nn\\
=& -\frac{1}{2} \ln( \det C) + \frac{1}{2} \ln (\det \uslash )
      + g\left( -\frac{1}{2}(\tr_{\mu} (C^2) - (\tr_{\mu} C)^2)   
          - \frac{1}{2} C_{\mu\nu} \tr_{\alpha}(\uslash \Gamma^{\mu}
      \uslash \Gamma^{\nu})  \right) \nn\\
& + g^2 \left(  \frac{3}{4}(\tr_{\mu}(C^4) - (\tr_{\mu} C^2)^2)
    + \frac{1}{4} C_{\mu\nu} C_{\rho\lambda} \tr_{\alpha}
    (\uslash \Gamma^{\mu}
      \uslash \Gamma^{\rho}\uslash \Gamma^{\nu}\uslash
      \Gamma^{\lambda})  \right) + {\cal O}(g^3) ,
\end{align}
where the solid and dashed lines represent fermion and boson
propagators, respectively, $\tr_{\mu}$ represents the trace in
the vector representation of SO(10), and $\tr_{\alpha}$ is the trace
taken in the left-handed spinor representation of SO(10).
In order to count the order, the formal coupling constant $g$ is
explicitly inserted.
We calculated this 2PI free energy to 7th order in $g$.
The numbers of graphs we computed are 2 at 0th, 1st and 
2nd orders, 4 at 3rd order, 12 at the 4th order and 49 at 5th order.
All of these were calculated in Ref. \cite{KKKMS1} and 
all the explicit graphs are presented in that paper.
Our new result is for 6th and 7th order.
The numbers of the graphs at 6th and 7th orders are 321 and
2346, respectively.
The generation and calculation of the planar 2PI graphs 
have been now totally automatized\footnote{%
In order to check the algorithm for generating graphs, we have used a
matrix $\phi^4$-QED model that provides the same 2PI graphs as those
of the IIB matrix model and is easier to compute analytically.
Then, we confirmed that the set of graphs and the symmetry factors 
are all correct.
(See Appendix \ref{sec:matrix_QED}.)}.

Next, we perform the Legendre transformation.
We define the variables conjugate to $V_i$ and $u$ as
\begin{align}
  M^i =& \frac{\partial}{\partial V_i} G(V,u), \\
  m   =& \frac{\partial}{\partial u} G(V,u) ,
\end{align}
and obtain the ordinary free energy via the Legendre transformation as 
follows:
\begin{equation}
  \label{eq:ord_F_IIB}
  F(M^i,m) = G - \sum_{i} M^i V_i - mu .
\end{equation}

Finally, the improved free energy $F_{\rm improved}$ to order $g^k$
can be obtained by subtraction, as in the case of the massless $\phi^4$
model discussed in \S \ref{sec:IMFA}:
\begin{equation}
  \label{eq:Fimp}
  F_{\rm improved}^k = \left. F(M^i - g M^i, m - gm) \right|_k .
\end{equation}
Here $|_k$ denotes expansion with respect to $g$, ignoring ${\cal
  O}(g^{k+1})$ terms and setting $g=1$.

Thus we obtain the improved free energies to 7th order.
As we have discussed, in general it is very difficult to identify
where plateaus are and how they grow.

In the next subsection, we find the extrema of these free energies at
several orders, and we compare their distribution with that for the $\phi^4$
matrix model.

\subsection{Extrema of the improved free energies and the extent of space-time}
\label{sec:extrema_Fimp}

We now have the improved free energies of the SO(4) and SO(7) ansatz
to 7th order in the improved perturbation.
In order to search for a plateau, we list all the extrema of these free
energies, as done in \S \ref{sec:IMFA}.
The extrema we have found are listed in Table \ref{tab:extrema_IIB}.

Now we consider the ``extent'' of space-time in two directions, 
which is defined by the moment of the eigenvalue distribution as follows:
\begin{align}
  R^2 & = \gf{\frac{1}{N} \; \Tr A_1^{\phantom{1}2} }
  = - \dder{F}{M^1}(M^i-gM^i , m-gm ) \ , \\
  r^2 & = \gf{\frac{1}{N} \; \Tr A_{10}^{\phantom{10}2} }
  = - \dder{F}{M^2}(M^i-gM^i , m-gm ).
\end{align}
We call $R$ the extent of ``our'' space-time and $r$ that of the internal one. 
The values of $R$, $r$ and their ratio, $\rho=R/r$, are also shown in Table
\ref{tab:extrema_IIB} for each extremum of the SO(4) and SO(7) ansatz.

\begin{table}[htbp]
  \begin{center}
    \leavevmode
    \begin{tabular}{c||c||l|l|l|l} \hline
      ansatz & order & \phantom{aaa} $F$ &  $\rho= R/r$ &
       \phantom{aa} $R^2$ & $\phantom{aa}r^2$ \\
       \hline \hline
    &  1st &\ph{$-$}5.52272  & 1.95530 & 0.410551 & 0.107383 \\ \cline{2-6}
    &  3rd &\ph{$-$}5.62672  & 1.91133 & 0.463883 & 0.126981 \\ \cline{3-6}
    &      &\ph{$-$}1.62094  & 2.15683 & 0.588327 & 0.12647 \\ \cline{2-6}
    &      &\ph{$-$}5.52146  & 1.92523 & 0.481755 & 0.129975 \\ \cline{3-6}
SO(7)&  5th &\ph{$-$}3.00676  & 2.19364 & 0.573357 & 0.11915 \\ \cline{3-6}
    &      &\ph{$-$}1.49243  & 2.0943  & 0.648315 & 0.147812 \\ \cline{2-6}
    &      &\ph{$-$}5.45127  & 1.93442 & 0.491098 & 0.13124 \\ \cline{3-6}
    &  7th &\ph{$-$}3.89291  & 2.39213 & 0.640411 & 0.111916 \\ \cline{3-6}
    &      &\ph{$-$}1.92312  & 1.62093 & 0.33974  & 0.129306 \\ \cline{3-6}
    &      &\ph{$-$}1.93939  & 1.93483 & 0.519378 & 0.138739 \\ \hline \hline
    &  1st &\ph{$-$}6.1533  & 1.85728 & 0.562580 & 0.163090 \\ \cline{2-6}
    &  3rd &\ph{$-$}6.34486  & 1.85336 & 0.650887 & 0.189489 \\ \cline{3-6}
    &      &\ph{$-$}0.696885 & 3.05943 & 1.34914  & 0.144137 \\ \cline{2-6}
    &  4th &\ph{$-$}1.17141  & 4.42435 & 1.90023  & 0.0970747 \\ \cline{3-6}
    &      &\ph{$-$}1.55668  & 2.58548 & 1.1568   & 0.173051 \\ \cline{2-6}
    &      &\ph{$-$}6.26112  & 1.92138 & 0.700007 & 0.189617 \\ \cline{3-6}
    &      &\ph{$-$}2.85746  & 5.00919 & 2.12045  & 0.0845069 \\ \cline{3-6}
SO(4)&  5th &\ph{$-$}0.5426   & 3.75008 & 1.81111  & 0.128784 \\ \cline{3-6}
    &      &\ph{$-$}0.0776919& 6.34998 & 2.92093  & 0.0724395 \\ \cline{3-6}
    &      &\ph{$-$}0.473299 & 3.31357 & 1.65917  & 0.151112 \\ \cline{2-6}
    &      & $-$2.62258 & 5.45278 & 2.69232  & 0.0905505 \\ \cline{3-6}
    &  6th & $-$2.78918 & 7.34015 & 3.55225  & 0.0659317 \\ \cline{3-6}
    &      & $-$0.642089& 7.36892 & 3.05248  & 0.056214 \\ \cline{3-6}
    &      &\ph{$-$}0.0100066& 3.50285 & 1.30668  & 0.106494 \\ \cline{2-6} 
    &      & $-$2.66095 & 3.11512 & 1.17719  & 0.377896  \\ \cline{3-6}
    &      &\ph{$-$}6.19216  & 1.97197 & 0.73242  & 0.188347 \\ \cline{3-6}
    &  7th &\ph{$-$}5.16286  & 6.42722 & 2.84737  & 0.0689285 \\ \cline{3-6}
    &      &\ph{$-$}2.75812  & 6.8741  & 2.22076  & 0.0469969 \\ \cline{3-6}
    &      &\ph{$-$}3.77441  & 9.26784 & 3.82752  & 0.0445615 \\ \hline
    \end{tabular}
    \caption{Extrema of the free energy and the extent of space-time
       for the SO(4)
       and SO(7) ansatz.}
    \label{tab:extrema_IIB}
  \end{center}
\end{table}

\begin{figure}[htbp]
  \begin{center}
    \leavevmode
  \begin{tabular}{c}
  \scalebox{1.0}{\includegraphics{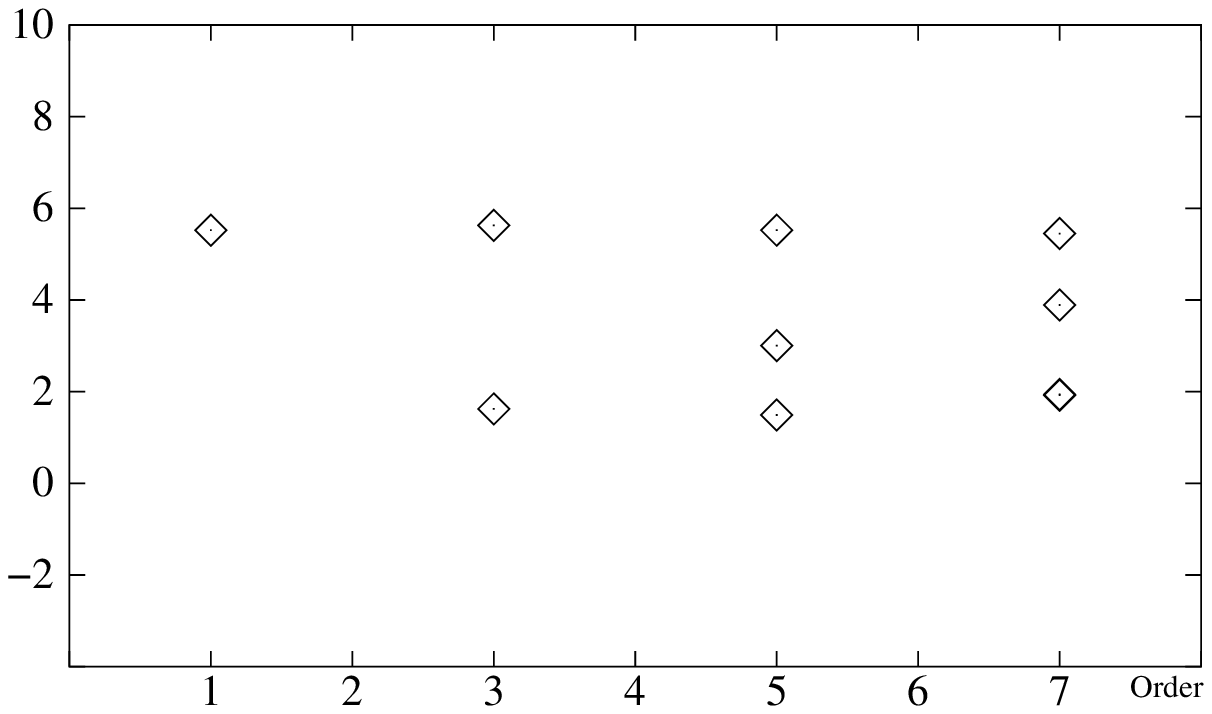}}  \\
  \scalebox{1.0}{\includegraphics{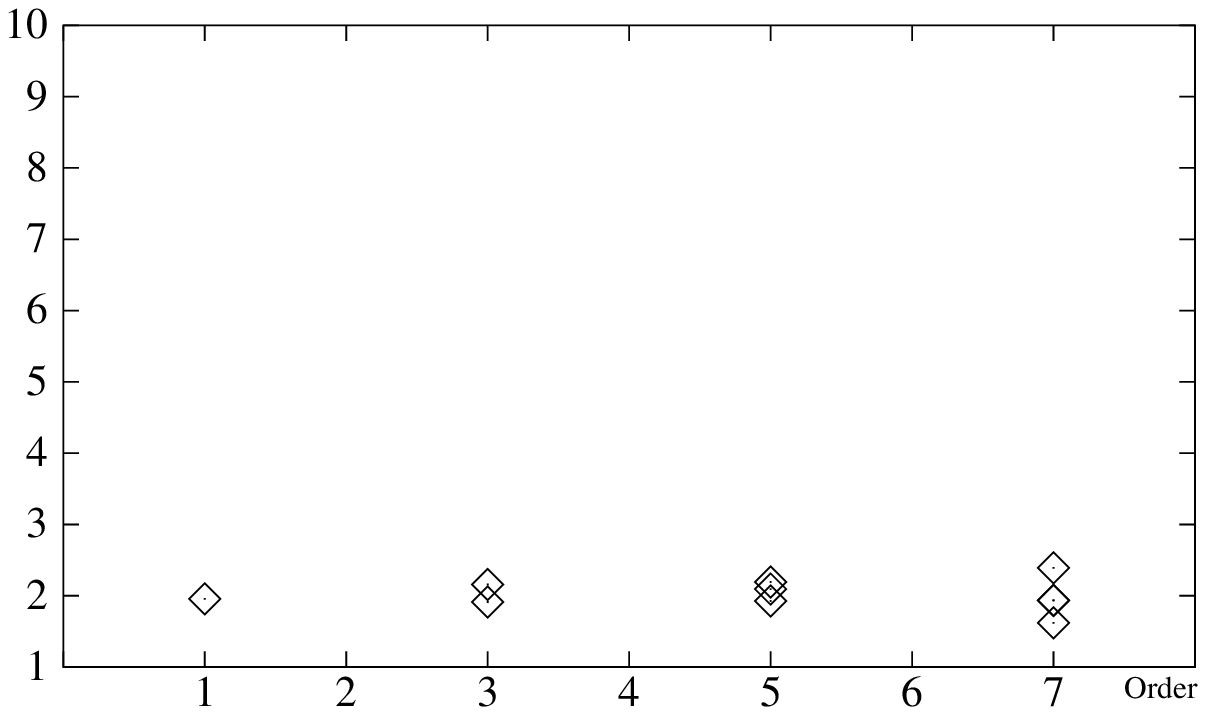}} \\
  \end{tabular}
  \captionwidth = 30em
    \hangcaption{Extrema of the free energy (upper) and ratio 
    of the extents (lower) for the SO(7) ansatz. }
    \label{fig:extrema}
  \end{center}
\end{figure}

\begin{figure}[htbp]
  \begin{center}
    \leavevmode
  \begin{tabular}{c}
  \scalebox{1.0}{\includegraphics{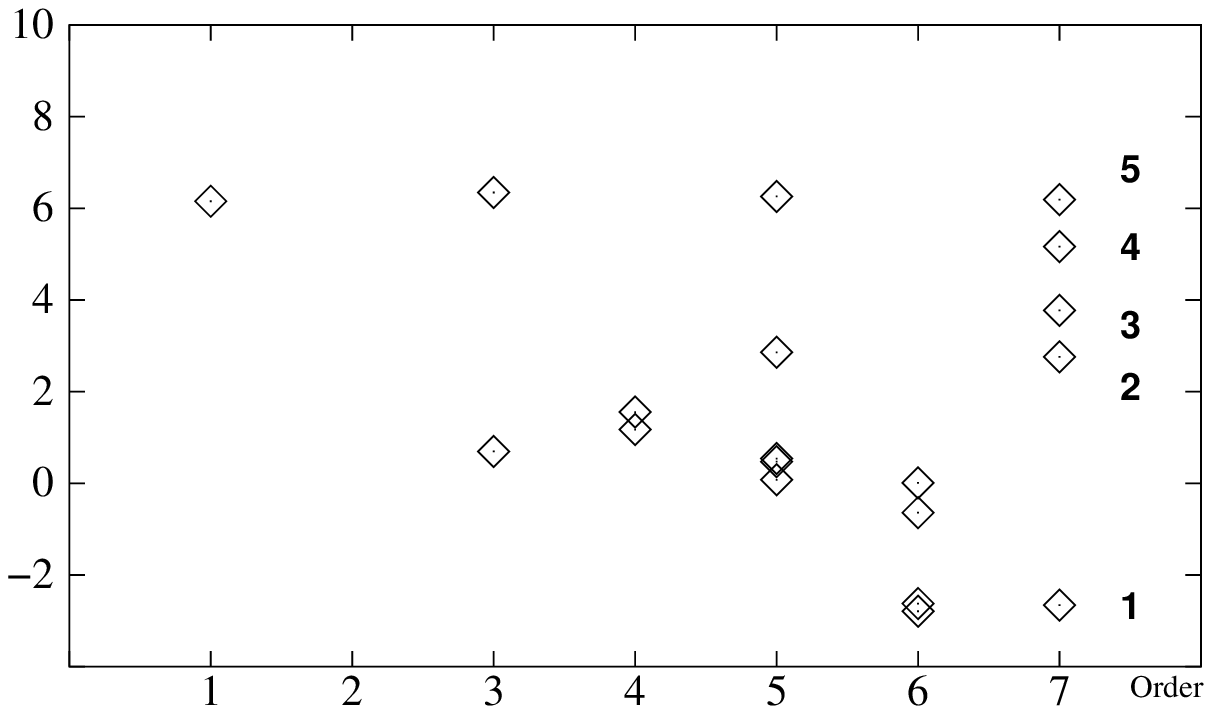}} \\
  \scalebox{1.0}{\includegraphics{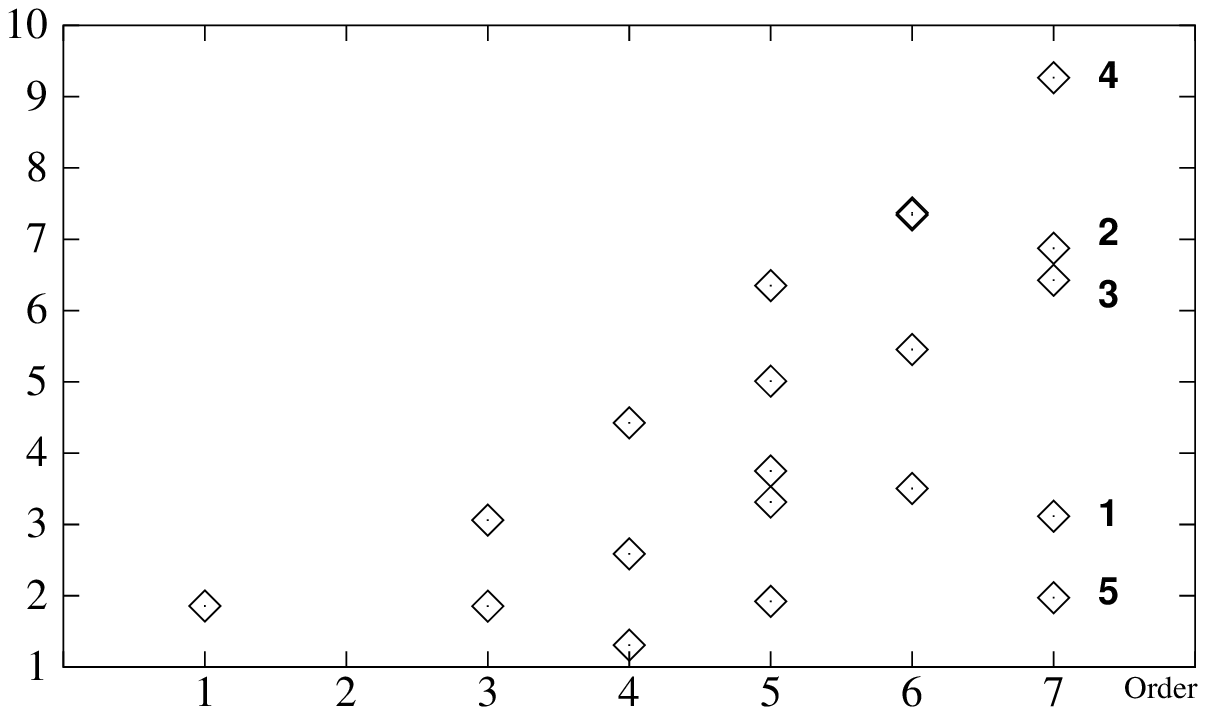}} \\
  \end{tabular}
  \captionwidth = 30em
    \hangcaption{Extrema of the free energy (upper) and ratio 
    of the extents (lower) for the SO(4) ansatz. 
    The numbers assigned at 7th order display a correspondence between 
    the extrema and the value of the ratio of the extents.}
    \label{fig:ratio}
  \end{center}
\end{figure}

In Figs. \ref{fig:extrema} and \ref{fig:ratio}, these
extrema and the ratios $\rho$ are plotted for each ansatz.

We find that these two ansatz yield significantly different behavior.
The SO(7) ansatz has fewer extrema than the SO(4) ansatz, and, 
in particular, it has no extrema at even orders.
Therefore, we speculate that extrema for the SO(7) ansatz do not accumulate
to form a plateau.
This means that the SO(7) ansatz is not a realistic assumption for the
eigenvalue distribution of the IIB matrix model, and thus a 
seven-dimensional flat space-time would not be realized as its stable vacuum.
Even if the SO(7) ansatz does develop a plateau at higher order, it would
not be regarded as a compactification, because the ratio of the extents
seems to be stabilized around 2.

In contrast, the SO(4) ansatz exhibits interesting behavior.
In this case, the number of extrema grows as the order increases.
It seems that even orders are not stable compared to odd orders.
We observe similar situations in various models.
Actually, in the zero-dimensional $\phi^4$ model, even lower orders
have no extrema, while odd orders develop a plateau even at lower
orders \cite{KKKMS1}.
Assuming this is the case, we ignore even orders.
Then we find peculiar behavior:
There are two characteristic extrema that can be taken as the
counterparts of the overshooting and undershooting extrema of the $\phi^4$
matrix model, and they are unrelated to a plateau.
It seems that the other extrema tend to accumulate and can be 
expected to form a plateau.
We further observe that the ratio of the extents for the SO(4) ansatz
are around 2 or 3 for these undershooting and overshooting extrema,
whereas the other extrema have rather large ratios, around 6 - 9.
This implies that on the plateau, the ratio of the extents takes a large
value.
In short, we can conclude that the SO(4) ansatz develops a plateau, and 
it predicts a quite large value for the ratio of the sizes of the internal and
external directions.
This indicates that our scenario for spontaneous compactification to
a flat four-dimensional space-time is promising.

\section{Conclusions and Discussions}
\label{sec:conlusion}

We have performed the improved mean field approximation for the IIB
matrix model up to 7th order and obtained the following conclusions:
\begin{itemize}
\item We first conclude that the SO(7) and SO(4) ansatz exhibit different
  types of behavior, as described below.
\item The SO(7) ansatz has fewer extrema than the SO(4) ansatz, and it 
  does not have a tendency to form a plateau.
  The eigenvalue distribution of the SO(7) ansatz is rather isotropic,
  and will not be realized as a compactification vacuum even if it has a 
  plateau.
\item The SO(4) ansatz has many extrema at higher orders and,
  except for the two special extrema mentioned below, it yields a quite
  large ratio of the extents of the four-dimensional space-time
  and that of the internal one.
\item The extrema of the SO(4) ansatz are distributed, as in the
  $\phi^4$ matrix model, which develops a plateau. The $\phi^4$
  matrix model has an overshooting and an undershooting extremum,
  which are located over and under the plateau, and the
  other extrema tend to accumulate.
  Thus, we conjecture that the IIB matrix model exhibits
  similar behavior and develops a plateau under the SO(4) ansatz.
\end{itemize}

At this stage we do not have clear plateaus for any ansatz, and
therefore we
cannot definitely tell which vacuum is realized in the IIB matrix model.
Indeed, in order to do this, we need to identify the plateau for each
ansatz, if such exists, and compare the values of the free energies 
at the plateaus.
If an ansatz has no plateaus, we conclude that it is not realized 
as a vacuum.
We expect that our SO(4) ansatz is close to the true vacuum, which
reproduces our universe, and it has a plateau where the free energy
has the lowest value.
In order to confirm this, we should analyze higher orders in the
improved perturbation series.
As mentioned above, our calculation is now totally automatized, 
and it seems possible to carry out further analysis with
the help of supercomputers.

\section*{Acknowledgments}
We would like to thank M. Nio for valuable discussions and useful
comments.
This work was supported in part by Grants-in-Aid for Scientific Research
from the Ministry of Education, Science, Sports and Culture of Japan
(\#1640290, \#03603, \#02846 and \#03529).
The work of T.~K., S.~K. and S.~S. is supported in part
by the Japan Society for the
Promotion of Science under the Post- and Pre-doctoral Research Program.

\appendix

\section{The 2PI free energy of $\phi^4$-QED type matrix model}
\label{sec:matrix_QED}

In this appendix we consider a matrix model that has a 
$\phi^4$ type and a QED type interaction term. The action is given by 
\begin{equation}
S=N\Tr\left(\frac{m_B}{2}A^2+\frac{g^2}{4}A^4+JA
           +\frac{m_F}{2}\psi^2+gA\psi^2
      \right),
\label{phi4QED1}
\end{equation}
where $A$ and $\psi$ are $N \times N$ matrices and $\psi$ 
is assumed to have flavor $f$. 
As we discuss below, we have introduced a source $J$ 
for $A$ in order to cancel the tadpole graphs. 

By comparing this model with the IIB matrix model with the mean field action 
(\ref{IIB_action_mod}), it is easy to see that (\ref{phi4QED1}) 
generates the same vacuum graphs with the same symmetry factors 
as the IIB matrix model.
These two models have the same types of propagators and vertices, 
except for the source term $JA$, which we explain below. 
Here $m_B$ and $m_F$ play the roles of $C^{-1}$ 
and $\uslash^{-1}$, respectively, and $f$ should be set to $-1$, because 
$\psi$ is fermionic in the IIB matrix model. However, we do not fix $f$, 
in order to classify graphs via the number of $\psi$-loops. 
It is worth noting that in (\ref{phi4QED1}) there is no symmetry 
that forces the one-point function $\gf{A}$ to vanish. 
This is in contrast to the situation in the IIB matrix model, 
where it cannot have a non-zero 
value, due to the Lorentz symmetry.
In order to eliminate unwanted one-point functions from our model, we
choose $J$ in such a way that $\gf{\Tr A}=0$ order by order in the
perturbation theory with respect to $g$. 

Our aim is to confirm that our list of the planar 2PI graphs 
of the IIB matrix model is complete.
For each graph, we read off the number of boson propagators (B), 
fermion propagators (F), fermion loops (L), $A^4$ vertices (V)
and $A\psi^2$ vertices (Y), and we deduce the 2PI free energy of
(\ref{phi4QED1}) as 
\begin{equation}
-\frac{1}{\mbox{symmetry factor}}\sum (-g^2)^{V}(-g)^Ym_B^{-B}m_F^{-F}f^{L}, 
\end{equation}
where the summation is taken over all possible planar 2PI graphs 
without tadpoles up to the given order of $g$. Note that in this expression 
cancellation between different graphs never occurs, 
because the symmetry factor is always positive. 
We compare this with the 2PI free energy of (\ref{phi4QED1}) 
in the large-$N$ limit, computed with a completely different method 
that we now explain. 

\subsection{Loop equation}
The field redefinition in (\ref{phi4QED1}) gives
\begin{equation}
S=N\Tr\left(\frac{1}{2}A^2+\frac{g^2}{4}A^4+JA
           +\frac{1}{2}\psi^2+\lambda gA\psi^2
      \right).
\label{phi4QED2}
\end{equation}
In order to recover the original parameters in (\ref{phi4QED1}), 
we only need to do is to make the replacements $g\rightarrow g/m_B,\ 
J\rightarrow J/\sqrt{m_B},\ \lambda\rightarrow \sqrt{m_B}/m_F$.

For the purpose of computing the 2PI free energy of (\ref{phi4QED2}), 
we first compute the two-point function $\gf{\Tr (A^2)}/N$ 
in the large-$N$ limit, 
retrieve the parameters $m_B$ and $m_F$, and integrate it 
with respect to $m_B$ to obtain the ordinary free energy. 
Finally, we carry out its Legendre transformation
in terms of $m_B$ and $m_f$ to obtain the 2PI free energy.

Performing the Gaussian integration in (\ref{phi4QED2}) 
with respect to $\psi$, we obtain 
\begin{equation}
S_{\rm eff}=N\Tr\left(\frac{1}{2}A^2+\frac{g^2}{4}A^4+JA\right)
             -\frac{f}{2}\sum_{p=1}^{\infty}\sum_{q=0}^{\infty}
             \frac{(-\lambda g)^{p+q}}{p+q}{}_{p+q}C_q\Tr(A^p)\Tr(A^q).
\end{equation}
In order to compute the two-point function $\gf{\Tr (A^2)}/N$ 
of this model, we start with the loop equation
(Schwinger-Dyson equation)
\begin{equation}
0=\int dA \frac{\partial}{\partial A^{\alpha}}
\Tr (A^n t^{\alpha})
         e^{-S_{\rm eff}},
\end{equation}
where $t^{\alpha}$ is the orthogonal basis of hermitian $N\times N$ 
matrices: $\Tr(t^{\alpha}t^{\beta})=\delta^{\alpha\beta}$.
By defining 
\begin{equation}
w_n=\frac{1}{N}\gf{\Tr (A^n)},
\end{equation}
this equation gives us the relation between these correlation
functions,
\begin{equation}
w_n=\sum_{m=0}^{n-2}w_mw_{n-2-m}-g^2w_{n+2}-Jw_{n-1}
       +f\sum_{p=1}^{\infty}\sum_{q=0}^{\infty}
             (-\lambda g)^{p+q}~{}_{p+q-1}C_qw_{n+p-2}w_q,
\label{loopeq1}
\end{equation}
where we have used the factorization property in the large-$N$
limit. Note that this equation holds even in the $n=1$ case 
if we ignore the first term on the right-hand side. 
The loop equation enables us to determine $w_n$ order by order 
in the perturbation theory as follows. 
First we expand each $w_n$ in terms of
$g$:
\begin{equation}
w_n=\sum_{k=0}^{\infty}g^kw_n^{(k)}.
\end{equation}
Substituting this into (\ref{loopeq1}) and comparing both sides 
to each order in $g$, we obtain the following equation:
\begin{eqnarray}
w_n^{(k)} & = & \sum_{m=0}^{n-2}\sum_{l=0}^{k}w_m^{(l)}w_{n-2-m}^{(k-l)}
                -w_{n+2}^{(k-2)}-Jw_{n-1}^{(k)} \nn \\
          & + & f\sum_{p=1}^{k}\sum_{q=0}^{k-p}\sum_{l=0}^{k-p-q}
                (-\lambda)^{p+q}~{}_{p+q-1}C_qw_{n+p-2}^{(l)}w_q^{(k-p-q-l)}.
\label{loopeq2}
\end{eqnarray}
Because $w_0=1/N\gf{\Tr 1}=1$, we have the ``boundary condition'' 
$w_0^{(i)}=\delta_0^i$.
Note that only the quantities $w_m^{(l)}$ with $l$ smaller 
than or equal to $k$ appear on the right-hand side and that 
when $w_m^{(k)}$ appears on the right-hand side, $m$ is always 
smaller than $n$.
Because of this property, (\ref{loopeq2}) together with the boundary
condition determines all the $w_n^{(k)}$. 

Once we determine $w_1^{(k)}$ up to a given order, for example, 
the 14th, corresponding to the 7th in the IIB matrix model, we can tune $J$ 
order by order in such a way that $w_1^{(k)}$ vanishes for each $k$. 
Substituting the resulting expression for $J$ into $w_2^{(k)}$, 
we can derive the two-point function up to the order we desire. 
Because of this procedure, graphs containing a tadpole as a
subgraph no longer contribute to $w_2^{(k)}$, and therefore 
the free energy obtained by integrating $w_2=\sum_{k}g^kw_2^{(k)}$ 
gives the sum of all the planar vacuum graphs without tadpoles. 
Once we have the free energy, it is easy to carry out the Legendre 
transformation to obtain the 2PI free energy.

We have compared the two forms of the 2PI free energy obtained 
using the two totally different 
methods and found complete agreement.
This proves that our list of all planar 
2PI graphs of the IIB matrix model is complete.



\begin{thebibliography}{99}


\bibitem{KKKMS1}
H.~Kawai, S.~Kawamoto, T.~Kuroki, T.~Matsuo and S.~Shinohara,
``Mean Field Approximation of IIB Matrix Model and Emergence of Four
Dimensional Space-Time,''
Nucl.\ Phys.\ B {\bf 647}, 153 (2002)
[arXiv:hep-th/0204240].

\bibitem{Nishimura:2001sx}
J.~Nishimura and F.~Sugino,
``Dynamical generation of four-dimensional space-time in the IIB
matrix model,''
JHEP {\bf 0205} 001, (2002)
[arXiv:hep-th/0111102]. \\
For a reference, see
D.~Kabat and G.~Lifschytz,
``Approximations for strongly-coupled supersymmetric quantum mechanics,''
Nucl.\ Phys.\ B {\bf 571}, 419 (2000)
[arXiv:hep-th/9910001].


\bibitem{CY_cpt}
For a review, \\
M.B.~Green, J.H.~Schwarz and E.~Witten,
``Superstring theory,'' vol.2
Cambridge monographs on mathematical physics
[ISBN:0521357535], \\
J.~Polchinski,
``String theory,'' vol.2
Cambridge monographs on mathematical physics
[ISBN:0521633044].


\bibitem{KLT1987}
H.~Kawai, D.~C.~Lewellen and S.~H.~Tye,
``Construction Of Fermionic String Models In Four-Dimensions,''
Nucl.\ Phys.\ B {\bf 288}, 1 (1987).


\bibitem{Matrices}
T.~Banks, W.~Fischler, S.~H.~Shenker and L.~Susskind,
``M theory as a matrix model: A conjecture,''
Phys.\ Rev.\ D {\bf 55}, 5112 (1997)
[arXiv:hep-th/9610043],

\bibitem{IIA_matrices}
L.~Motl,
``Proposals on nonperturbative superstring interactions,''
[arXiv:hep-th/9701025], \\
T.~Banks and N.~Seiberg,
``Strings from matrices,''
Nucl.\ Phys.\ B {\bf 497}, 41 (1997)
[arXiv:hep-th/9702187], \\
R.~Dijkgraaf, E.~Verlinde and H.~Verlinde,
``Matrix string theory,''
Nucl.\ Phys.\ B {\bf 500}, 43 (1997)
[arXiv:hep-th/9703030].

\bibitem{typeI_matrices}
H.~Itoyama and A.~Tokura,
``USp(2k) matrix model: F theory connection,''
Prog.\ Theor.\ Phys.\  {\bf 99}, 129 (1998)
[arXiv:hep-th/9708123], \\
H.~Itoyama and A.~Tsuchiya,
Prog.\ Theor.\ Phys.\ Suppl.\  {\bf 134}, 18 (1999)
[arXiv:hep-th/9904018].

\bibitem{heterotic_matrices}
S.~J.~Rey,
``Heterotic M(atrix) strings and their interactions,''
Nucl.\ Phys.\ B {\bf 502}, 170 (1997)
[arXiv:hep-th/9704158], \\
D.~A.~Lowe,
``Heterotic matrix string theory,''
Phys.\ Lett.\ B {\bf 403}, 243 (1997)
[arXiv:hep-th/9704041].

\bibitem{Ishibashi:1997xs}
N.~Ishibashi, H.~Kawai, Y.~Kitazawa and A.~Tsuchiya,
``A large-N reduced model as superstring,''
Nucl.\ Phys.\ B {\bf 498}, 467 (1997)
[arXiv:hep-th/9612115].


\bibitem{new_matrices}
L.~Smolin,
``M theory as a matrix extension of Chern-Simons theory,''
Nucl.\ Phys.\ B {\bf 591}, 227 (2000)
[arXiv:hep-th/0002009], \\
T.~Azuma, S.~Iso, H.~Kawai and Y.~Ohwashi,
``Supermatrix models,''
Nucl.\ Phys.\ B {\bf 610}, 251 (2001)
[arXiv:hep-th/0102168], \\
L.~Smolin,
``The exceptional Jordan algebra and the matrix string,''
[arXiv:hep-th/0104050], \\
V.~Periwal,
``Matrices on a point as the theory of everything,''
Phys.\ Rev.\ D {\bf 55}, 1711 (1997)
[arXiv:hep-th/9611103], \\
T.~Yoneya,
``Schild action and space-time uncertainty principle in string theory,''
Prog.\ Theor.\ Phys.\  {\bf 97}, 949 (1997)
[arXiv:hep-th/9703078], \\
A.~Fayyazuddin, Y.~Makeenko, P.~Olesen, D.~J.~Smith and K.~Zarembo,
``Towards a non-perturbative formulation of IIB superstrings by matrix
models,''
Nucl.\ Phys.\ B {\bf 499}, 159 (1997)
[arXiv:hep-th/9703038], \\
N.~Kitsunezaki and J.~Nishimura,
``Unitary IIB matrix model and the dynamical generation of the space
time,''
Nucl.\ Phys.\ B {\bf 526}, 351 (1998)
[arXiv:hep-th/9707162], \\
S.~Hirano and M.~Kato,
``Topological matrix model,''
Prog.\ Theor.\ Phys.\  {\bf 98}, 1371 (1997)
[arXiv:hep-th/9708039], \\
T.~Tada and A.~Tsuchiya,
``Toward a supersymmetric unitary matrix formulation of the IIB matrix
model,''
Prog.\ Theor.\ Phys.\  {\bf 103}, 1069 (2000)
[arXiv:hep-th/9903037].


\bibitem{Aoki:1999bq}
H.~Aoki, S.~Iso, H.~Kawai, Y.~Kitazawa, A.~Tsuchiya and T.~Tada,
``IIB matrix model,''
Prog.\ Theor.\ Phys.\ Suppl.\  {\bf 134} (1999) 47
[arXiv:hep-th/9908038].


\bibitem{Sugino:2001}
S.~Oda and F.~Sugino,
``Gaussian and mean field approximations for reduced Yang-Mills integrals,''
JHEP {\bf 0103} 026, (2001)
[arXiv:hep-th/0011175];\\
F.~Sugino,
``Gaussian and mean field approximations for reduced 4D supersymmetric
Yang-Mills integral,''
JHEP {\bf 0107} 014, (2001)
[arXiv:hep-th/0105284].

\bibitem{Nishimura-Okubo-Sugino}
J.~Nishimura, T.~Okubo and F.~Sugino,
``Convergent Gaussian expansion method: Demonstration in reduced  Yang-Mills integrals,''
JHEP {\bf 0210}, 043 (2002)
[arXiv:hep-th/0205253].

\bibitem{Stevenson:1981vj}
P.~M.~Stevenson,
``Optimized Perturbation Theory,''
Phys.\ Rev.\ D {\bf 23}, 2916 (1981), \\
A.~Dhar,
``Renormalization Scheme - Invariant Perturbation Theory,''
Phys.\ Lett.\ B {\bf 128}, 407 (1983).

\bibitem{Brezin:1978sv}
E.~Brezin, C.~Itzykson, G.~Parisi and J.~B.~Zuber,
``Planar Diagrams,''
Commun.\ Math.\ Phys.\  {\bf 59}, 35 (1978).

\bibitem{Nishimura:2000}
J. Nishimura and G. Vernizzi,
``Spontaneous breakdown of Lorentz invariance in IIB matrix model,''
JHEP {\bf 0004} 015, (2000)
[arXiv: hep-th/0003223]; \\
``Brane world from IIB matrices,''
Phys.\ Rev.\ Lett.\ {\bf 85}, 4664 (2000)
[arXiv: hep-th/0007022]; \\
J. Ambjorn, K.N. Anagnostopoulos, W. Bietenholz, T. Hotta and J. Nishimura,
``Monte Carlo studies of the IIB matrix model at large N,''
JHEP {\bf 0007} 011, (2000)
[arXiv: hep-th/0005147]; \\
K.N. Anagnostopoulos and J. Nishimura,
``A new solution of the complex action problem and the dynamical
space-time in the IIB matrix model,''
[arXiv: hep-th/0108041];\\
Jun Nishimura,
``Exactly Solvable Matrix Models for the Dynamical Generation of
Space-Time in Superstring Theory,''
Phys.\ Rev.\ D {\bf 65}, 105012 (2002)
[arXiv: hep-th/0108070].



\end{thebibliography}
\end{document}